\documentclass[conference]{IEEEtran}
\IEEEoverridecommandlockouts
\usepackage{cite}
\usepackage{amsmath,amssymb,amsfonts}
\usepackage{algorithmic}
\usepackage{graphicx}
\usepackage{textcomp}
\usepackage{xcolor}
\def\BibTeX{{\rm B\kern-.05em{\sc i\kern-.025em b}\kern-.08em
    T\kern-.1667em\lower.7ex\hbox{E}\kern-.125emX}}
    
\usepackage{tikz}
\usepackage{pgfplots}
\usetikzlibrary{spy}

\DeclareMathOperator*{\argmax}{arg\,max}

\def\*#1{\mathbf{#1}}
\newcommand{\vect}[1]{{\mathbf{#1}}}
\newcommand{\mat}[1]{{\mathbf{#1}}}

\renewcommand{\sf}[1]{\mathsf{#1}}
\renewcommand{\intercal}{{T}}

\newcommand{\SDoF}{\textnormal{Sum-DoF}}
\newcommand{\SDoFlb}{\textnormal{Sum-DoF}_\textnormal{Lb}}
\newcommand{\modd}{\textnormal{ } \mathsf{mod}_1 \textnormal{ } }
\renewcommand{\mod}{ \textnormal{ } \mathsf{mod} \textnormal{ } }

\newcommand{\yue}[1]{{\color{brown}#1}}

\newcommand{\red}[1]{{\color{red}#1}}
\newcommand{\bl}[1]{{\color{blue}#1}}

\newcommand{\lo}[1]{}
\newcommand{\sh}[1]{{#1}}
\newif\ifADDpagenumber
\ADDpagenumberfalse

\allowdisplaybreaks[4]
\usepackage{amsthm}
\newtheorem{theorem}{Theorem}
\newtheorem{definition}{Definition}
\newtheorem{lemma}{Lemma}
\newtheorem{corollary}{Corollary}
\newtheorem{property}{Property}

\newtheorem{example}{Example}

\begin{document}

\title{DoF Analysis for (M, N)-Channels through a Number-Filling Puzzle}

 \author{
 \IEEEauthorblockN{Yue Bi}
 \IEEEauthorblockA{\textit{LTCI, Telecom Paris, IP Paris} \\
 91120 Palaiseau, France\\
 yue.bi@telecom-paris.fr}
 \\
 \IEEEauthorblockN{Philippe Ciblat}
 \IEEEauthorblockA{\textit{LTCI, Telecom Paris, IP Paris} \\
 91120 Palaiseau, France\\
 philippe.ciblat@telecom-paris.fr}
 \and
 \IEEEauthorblockN{Mich\`ele Wigger}
 \IEEEauthorblockA{\textit{LTCI, Telecom Paris, IP Paris} \\
 91120 Palaiseau, France\\
 michele.wigger@telecom-paris.fr}
 \\
 \IEEEauthorblockN{Yue Wu}
 \IEEEauthorblockA{\textit{School of Cyber Science and Engineering,}\\ Shanghai Jiao Tong University, China \\
 wuyue@sjtu.edu.cn}
 }

\author{
    \IEEEauthorblockN{Yue Bi\IEEEauthorrefmark{1}\IEEEauthorrefmark{2} , Yue Wu\IEEEauthorrefmark{1}, Cunqing Hua\IEEEauthorrefmark{1}}
    \IEEEauthorblockA{\IEEEauthorrefmark{1}\textit{School of Electronic Information and Electrical Engineering, Shanghai Jiao Tong University}, China\\ 
   	\{wuyue, cqhua\}@sjtu.edu.cn }
    \IEEEauthorblockA{\IEEEauthorrefmark{2} \textit{LTCI, Telecom Paris, IP Paris}, 91120 Palaiseau, France,   bi@telecom-paris.fr}
}

\maketitle

\ifADDpagenumber
\thispagestyle{plain} % add page number
\pagestyle{plain}
\fi

\begin{abstract}
%THIS PAPER IS ELIGIBLE FOR THE STUDENT PAPER AWARD.
We consider a $\sf K$ user  interference network with general connectivity, described by a matrix $\mat{N}$,  and general message  flows, described by  a matrix $\mat{M}$. Previous studies have demonstrated that the standard interference scheme (IA) might not be optimal for networks with sparse connectivity.  In this paper, we formalize   a general IA coding scheme and  an intuitive number-filling puzzle for given $\mat{M}$ and $\mat{N}$ in a way that the score of the solution to the puzzle determines the optimum sum degrees that can be achieved by the IA scheme. A solution to the puzzle is proposed  for a general  class of symmetric channels, and it is shown that this solution leads to significantly higher  $\SDoF$ than the standard IA scheme.
\end{abstract}


\section{Introduction}
The sum degrees of freedom ($\SDoF$) characterizes the pre-log approximation of the sum-capacity in the asymptotic regime of infinite Signal-to-Noise Ratios (SNR)  \cite{Algoet}, i.e., when the network operates in the interference-limited regime. Identifying the achievable region of $\SDoF$ for a multi-user channel is an important topic. The study of the $\SDoF$ of interference channels (IC) and X-channels, has a rich history, see e.g., \cite{foschini1998limits, lapidoth_cognitive_2007, devroye2007multiplexing, cadambe_interference_2008, jafar_degrees_2008,cadambe_interference_2009,annapureddy_degrees_2012,cadambe_can_2008, motahari_real_2014, zamanighomi_degrees_2016, wei_iterative_2019}. Notably, it has been demonstrated that the $\SDoF$ of a fully-connected $\sf K$-user IC without cooperation is $\sf K/2$ when the channel coefficients are independent and identically distributed (i.i.d.) fading according to a continuous distribution \cite{jafar_degrees_2008}. Furthermore, the $\SDoF$ of the corresponding X-channel is ${\sf K}^2/(2{\sf K}-1)$. Both of these $\SDoF$s are achieved through interference alignment (IA) schemes \cite{cadambe_can_2008}.   %For general networks and message flows the optimal assignment of precoding matrices to the various messages are unknown to date. In this article we formalize the $\SDoF$ that can be achieved with a given a  important aspect of the design of the IA scheme is the allocation of precoding matrices. On one hand, this allocation must guarantee that useful messages are not subject to interference from other messages. On the other hand, the assignment should ensure that the interference space remains small in comparison to the overall signal space. For X-channel, authors of \cite{jafar_degrees_2008} assign the same precoding matrix for messages intended for the same Rx. This allocation is proven to be optimal for the X-channel.
 In our previous work in \cite{isit2022_bi}, we showed that this idea of  assigning  precoding matrices to the messages achieves only a suboptimal $\SDoF$ for a specific non-fully connected network. % where  and better  performance can be achieved by reducing %of a partially connected X-channel where each Rx observes a linear combination of all Tx-signals in Gaussian noise, except for the signal sent by its corresponding Tx. For this channel, we achieved an improvement in $\SDoF$  by reducing 
 the number of precoding matrices and rearranging them. %, which underscores the observation that the optimal precoding matrix allocation method for X-channel may not be optimal for other channels, especially those with sparse connectivity.
%Motivated by these findings, in this paper, we study the
In this paper we generalize our previous findings  for general network connectivities and general message flows by formalizing an IA scheme and a number-finding puzzle whose solution determines the choice of the assigned precoding matrices, and thus of the   $\SDoF$ achieved by the IA scheme. 

More formally, we consider networks with $\sf K$ transmitters and $\sf K$ receivers. The network connectivity is characterized by a matrix $\mat{N} \in \{0, 1\}^{(\sf K \times \sf K)}$, where the entry in row-$p$ and column $q$ equals $1$ if the signal sent by Transmitter (Tx) $q$ interferes at Receiver (Rx)  $p$, see \eqref{eq:gen}. The message flow is described by a  matrix $\mat{M} \in \{0, 1\}^{(\sf K \times \sf K)}$, where the entry in row-$p$ and column $q$ equals $1$ if Tx $q$ sends a message to Rx $p$. %  In a given channel, each Rx observes a noisy linear combination of signals from Txs where the corresponding entries of matrix $\mat{N}$ equal to $1$, and each Tx only sends a message to Rxs where the corresponding entries of matrix $\mat{M}$ equal to $1$. 
For example, the following matrices represent a standard 3-user interference channel
\begin{equation} \label{eqn:intro_example}
	\mat{M} = \begin{bmatrix}
		1 & 0 & 0\\
		0 & 1 & 0 \\
		0 & 0 & 1
	\end{bmatrix}, \hspace{1cm}
	\mat{N} = \begin{bmatrix}
		1 & 1 & 1 \\
		1 & 1 & 1 \\
		1 & 1 & 1 
	\end{bmatrix},
\end{equation}
where each Tx $q \in \{1,2,3\}$ sends a message  only to its corresponding Rx, and each Rx observes a linear combination of all the transmit  signals corrupted in additive white Gaussian noise.

\textit{The puzzle:} We present the number-filling puzzle for given connectivity and message flow matrices $\mat{M}$ and $\mat{N}$. 
%We imply a matrix $\mat{G}$ to represent the allocation of the precoding matrix. Each entry of $\mat{G}$ indicates the index of the precoding matrix applied to the corresponding message. Designing a new IA scheme is a challenging task, hence we formalize an intuitive number-filling puzzle to find an improved precoding matrix allocation. In the puzzle, $\mat{M}$ and $\mat{N}$ are given and participants need to fill an matrix 

Fill in the entries of a matrix $\sf K$-by-$\sf K$ matrix 
$\mat{G}$ with non-negative integers, i.e., elements of $\mathbb{N}_0$, according to the two rules
\begin{itemize}
%\item ($\mat{G}$ has non-negative integers as entries. I.e.,  $\mat{G}[p,q] \in \mathbb{Z}_0^+$, where  denotes the set of non-negative integers. 
\item  $\mat{G}$ inherits zeros from $\mat{M}$. I.e., $\mat{M}[p,q]=0 \Rightarrow \mat{G}[p,q]=0$. 
	\item All non-zero entries   in a column are different. I.e., $\mat{G}[p,q]= \mat{G}[p',q]$ $\Rightarrow$ $\mat{G}[p,q]=0$. 
\end{itemize}
The goal of the puzzle is to maximize the following score 
\begin{equation}\label{eq:score}
	\mathsf{S}=\frac{  \|\mat{G}\|_0}{ \max_{p \in [\sf K]}\left\{ \|\mat{G}[p,:]\|_0 + g^{(p)} \right\}} 
\end{equation}
where $\|\mat{G}[p,:]\|_0$ denotes the number  of non-zero entries in the $p$-th row of $\mat{G}$ and $g^{(p)}$ denotes the number of different non-zeros integers in the submatrix $\mat{G}^{(p)}$, which is obtained from $\mat{G}$ by removing the  $p$-th row  as  well as all the  columns  $q $ for which the entry $\mat{N}[p,q] = 0$. 

%After filling the matrix, the participant needs to determine two numbers for each $p \in \{1, 2, \cdots \sf K\}$: $\|\mat{M}[p,:]\|_0$, representing the number of non-zeros entries in the $p$-th row of $\mat{M}$; $g^{(p)}$, representing the number of different non-zeros integers in the sub-matrix $\mat{G}^{(p)}$ obtained by removing certain rows and column of $\mat{G}$ based on the value of $p$ (details are provided in Section \ref{sec:main}). The score of the filling method is given by 

%The objective of this puzzle is to find the filling method that \emph{minimize} the score.
\begin{example}For the connectivity and message flow matrices in \eqref{eqn:intro_example}, two possible filling could be
\begin{equation} \label{eqn:intro_example_pre}
	\mat{G} = \begin{bmatrix}
		1 & 0 & 0\\
		0 & 1 & 0\\
		0 & 0 & 1
	\end{bmatrix} \text{ or }
	\mat{G} = \begin{bmatrix}
		1 & 0 & 0\\
		0 & 2 & 0\\
		0 & 0 & 3
	\end{bmatrix}.
\end{equation}
In this case, the submatrix $\mat{G}^{(p)}$ is obtained by removing the $p$-th row of $\mat{G}$. It is evident that the first filling yields a better score as $g^{(p)} = 1$ for all values of $p$ in the first filling, whereas $g^{(p)} = 2$ for all values of $p$ in  the second filling. Since  in both fillings $ \|\mat{G}\|_0 =3$  and $ \|\mat{G}[p,:]\|_0 =1$ for each $p$, we obtain for  the first filling the score $\mathsf{S}= \frac{ 3}{2}$, which is exactly the {\SDoF} of the 3-user interference channel. (The second filling would have yield $\mathsf{S}= \frac{ 3}{3}=1$.)
\end{example} 

\begin{example} 
We look at an example with the matrix $\mat{M}$ and $\mat{N}$ are given below. 
\begin{equation}\label{eqn:example_mm}
	\mat{M} = \mat{N} = 
	\begin{bmatrix}
		0 & 1 & 1 & 1 \\
		1 & 0 & 1 & 1 \\
		1 & 1 & 0 & 1 \\
		1 & 1 & 1 & 0 
	\end{bmatrix}.
\end{equation}
This channel is similar to a X-channel except there is one less link for each Rx. Two possible filling are
\begin{equation}
	\mat{G}=
	\begin{bmatrix}
		0 & 1 & 1 & 1 \\
		2 & 0 & 2 & 2 \\
		3 & 3 & 0 & 3 \\
		\red{0} & 2 & 3 & 0 
	\end{bmatrix} \text{ or }
	\mat{G}=
	\begin{bmatrix}
	0 & 1 & 1 & 1 \\
	2 & 0 & 2 & 2 \\
	3 & 3 & 0 & 3 \\
	4 & 4 & 4 & 0 
	\end{bmatrix}
\end{equation}
In this case, $\mat{G}^{(p)}$ is obtain by removing the $p$-th row and $p$-th column in $\mat{G}$. We first examine the filling on the left. We find that $g^{(p)} = 2$ for $p \in \{1,2,3\}$ as the integer $p$ is not present in $\mat{G}^{(p)}$, whereas $g^{(4)} = 3$. Notice that $\mat{G}$ on the left has one less positive integer compared to the message flow matrix $\mat{M}$ (marked in red), so $\|\mat{G}[p,:]\|_0 = 3$ for $p \in \{1,2,3\}$ and $\|\mat{G}[4,:]\|_0 = 2$, which yields the score $\sf S = \frac{11}{5}$. For the filling on the right, we have $\|\mat{G}[p,:]\|_0 = 3$ and $g^{(p)} = 3$ for every rows. We obtain the score $\sf S = 2$ which is smaller than the score first filling. It is intriguing to discover that a higher score can be achieved by reducing the total number of positive integers in $\mat{G}$, i.e.$\|\mat{G}\|_0$, providing valuable inspiration for this work.
\end{example} 

In Theorem \ref{thm:dof_bounds} of Section \ref{sec:main}, we demonstrate that \eqref{eq:score} provides the  $\SDoF$ that is achievable through IA for the interference network with connectivity matrix $\mat{N}$ and message flow matrix $\mat{M}$. To prove the theorem, we construct an IA scheme that assigns the  precoding  matrices as described by the solution  $\mat{G}$ of the puzzle.  In this construction, the dimension of the signal space at Rx $p$ is proportional to $\|\mat{G}[p,:]\|_0$  (where the second rule in the puzzle ensures that signals spaces corresponding to different messages are disjoint), while the dimension of its interference space is proportional to $g^{(p)}$. 

%
%n Section \ref{sec:proof_cor1} the puzzle is applied to find new IA scheme symmetric channels, which achieves an improved $\SDoF$ compared to the naive implementation of the scheme for X-channel.
%

\textit{Notations:} We use sans serif font for constants, bold for vectors and matrices, and calligraphic font for most sets. The sets of complex numbers and nonnegative integers are denoted $\mathbb{C}$ and $\mathbb{N}_0$.  For a finite set $\mathcal{A}$, let $|\mathcal{A}|$ denote its cardinality. For any $n\in\mathbb{Z}^+$,  define $[n] \triangleq \{1,2,\ldots,  n\}$. The operation $\modd$ represents modulo with an offset of 1, equivalent to regular modulo $\mod$ for a non-zero remainder, and it returns the divisor if the remainder is zero. For any vector $\vect{v}$, let $\textnormal{diag}(\vect{v})$ be the diagonal matrix with diagonal entries given by the elements of the vector $\vect{v}$. 
When writing $[ \boldsymbol{v}_i\colon i \in \mathcal{S}]$ or $[\boldsymbol{v}_i]_{i\in \mathcal{S}}$ we mean the matrix consisting of the set of columns $\{\vect{v}_i\}_{i\in\mathcal{S}}$. For a matrix $\mat{A} \in \mathbb{R}^{(n\times n)}$ and finite sets $\mathcal{A}$, $\mathcal{B} \subseteq \mathbb{Z}$ , we denote $\mat{A}[\mathcal{A}, \mathcal{B}]$ as the submatrix of $\mat{A}$ obtained by removing rows in $[n]\backslash (\mathcal{A} \modd n)$ and columns in $[n]\backslash (\mathcal{B} \modd n)$. Let further $\mat{A}[\mathcal{A}, :] \triangleq \mat{A}[\mathcal{A}, [n]]$, and $\|\mat{A}\|_0$ denote the number of non-zeros elements in $\mat{A}$. For two matrices $\mat{A}$, $\mat{B} \in \{0,1\}^{(n\times n)}$, the relationship $ \mat{A} \subseteq \mat{B} $ holds if and only if $\mat{A}$ can be obtained by replacing certain entries in $\mat{B}$ with $0$.
%We abbreviate \emph{independent and identically distributed} by \emph{i.i.d.}.

\section{Channel model}\label{sec:channel_model}
Consider a network with $\sf  K$ Txs and $\sf K$ Rxs labeled from $1$ to $\sf K$. The connectivity of the network is described by the matrix $\mat{N} \in \{0, 1\}^{(\sf K \times \sf K)}$, and the message transmission is described by the matrix $\mat{M} \in \{0, 1\}^{(\sf K \times \sf K)}$. For simplicity, we call this network a $(\mat{M}, \mat{N})$-channel.
In this channel, each Rx $p$ observes a linear combination of the signals sent by the Txs with $\mat{N}[p,q]=1$ corrupted by  Gaussian noise.  Denoting Tx~$q$'s slot-$t$ input by $X_{q}(t) \in \mathbb{C}$ and Rx~$p$'s slot-$t$ output by $Y_{p}(t) \in \mathbb{C}$, the input-output relation of the network is expressed as:
\begin{IEEEeqnarray}{rCl}\label{eq:gen}
	Y_p(t) = \sum_{ \{q \colon \mat{N}[p,q]=1 \}} H_{p,q}(t) X_{q}(t) + Z_{p}(t) , \qquad p \in [\sf K], \IEEEeqnarraynumspace
\end{IEEEeqnarray}
where the sequences of complex-valued channel coefficients $\{H_{p,q}(t)\}$  and standard circularly symmetric Gaussian noises $\{Z_{p}(t)\}$ are both i.i.d. and  independent of each other and of all other channel coefficients and noises. 
Each coefficient $H_{p,q}(t)$ has independent and identically distributed real and imaginary parts, following a specified continuous distribution over a bounded interval $[-\sf H_{\max}, \sf H_{\max}]$. Importantly, these coefficients are known to all terminals even before communication commences.

The message flow in this channel is represented by the matrix $\mat{M}$, in the sense that   Tx $q$  transmits an independent message $a_{p,q}$ to each Rx~$p$ for which $\mat{M}[p,q]=1$. When communication is of blocklength $\sf T$, each message is uniformly distributed over $\left[2^{\sf T\sf R_{p,q}}\right]$, where $\sf R_{p,q}\geq 0$ denotes the rate of transmission, and it is independent of all other messages and of all channel coefficients and noise sequences. 
As a consequence, Tx~$q\in[\sf K]$ produces its block of channel inputs $\vect{X}_q \triangleq (X_{q}(1), \ldots, X_q(\sf T))$ as 
\begin{equation}
	\vect{X}_q = f_q^{(\sf T)}\left(\left\{a_{p,q} \colon \mat{M}[p,q]=1\right\}\right)
\end{equation}
by means of an encoding function $f_q^{(\sf T)}$ on appropriate domains and so that the  inputs satisfy the  block-power constraint
\begin{IEEEeqnarray}{rCl}\label{eq:power}
	\frac{1}{\sf T} \sum_{t=1}^{\sf T}     \mathbb{E}\left[|X_{q}(t)|^2\right] \leq\sf  P, \qquad q\in[\sf K].
\end{IEEEeqnarray}

Given a power $\sf P>0$, the \emph{capacity region} $\mathcal{C}(\sf P)$ is defined as the set of all  rate tuples $(\sf R_{p,q} \colon p, q \in [\sf K], \; \mat{M}[p,q]=1 )$ so that  for each blocklength $\sf T$  there exist
encoding functions $\{f_{q}^{(\sf T)}\}_{q \in [\sf K]}$ as described above and decoding functions $\{g_{p,q}^{(\sf T)}\}$ on appropriate domains producing the estimates 
\begin{equation}
	\hat{a}_{p,q} = g_{p,q}^{(\sf T)}( Y_p(1),\ldots, Y_p(\sf T)), \quad p, q\in[\sf K], \; \mat{M}[p,q] = 1,
\end{equation}
in a way that the sequence of error probabilities 
\begin{equation} \label{eq:error_prob}
	p^{(\sf T)}(\textnormal{error}) \triangleq  \textnormal{Pr}\bigg[ \bigcup_{ \substack{p\in[\sf K]}} \bigcup_{ \{ q \colon \mat{M}[p,q]=1\} } \hat{a}_{p,q}\neq a_{p,q} \bigg]
\end{equation}
tends to 0 as the blocklength $\sf T \to \infty$.
%\begin{IEEEeqnarray}{rCl}
%R_{max}(\sf P) = \max_{\vect{R} \in \mathcal{C}} \left( \sum_{p=1}^K \sum_{k=1, }^{\tilde{K}} R_p^{(k)}\right)
%\end{IEEEeqnarray}

Our main interest is in the \SDoF:%,% which characterizes the logarithmic growth in the high power regime of the maximum sum of all rates  tuples inside the capacity region:
\begin{IEEEeqnarray}{rCl}
	\SDoF \triangleq \varlimsup_{\sf P \rightarrow \infty} \sup_{ \vect{R} \in \mathcal{C}(\sf P)} \sum_{(p, q) \colon \mat{M}[p,q]=1} \frac{  \sf R_{p,q}}{\log \sf P}.
\end{IEEEeqnarray}

\section{Main Results}\label{sec:main}
The main results of this paper are lower bounds on the {\SDoF} of the network described in the previous Section~\ref{sec:channel_model}. To comprehend these results, we first present the definition of a valid precoding index matrix.
%\texttt{\color{red}$\mat{G}$ only depends on $\mat{M}$ and not on $\mat{N}$.}
\begin{definition}\label{def:valide_precoding_index_matrix}
	For a given message matrix $\mat{M} \in \{0, 1\}^{\sf K \times \sf K}$, a matrix $\mat{G}\in \mathbb{N}_0^{\sf K \times \sf K}$ is called a \textbf{valid precoding index matrix} if the following two requirements are satisfied:
	\begin{itemize}
		\item  If $\mat{M}[p, q] = 0$ then $\mat{G}[p, q] = 0$; 
		\item If $\mat{G}[p, q]= \mat{G}[p', q]$ for $p\neq p'$, then $\mat{G}[p, q] = 0$ 
	\end{itemize}
\end{definition}

Also denote by $\mat{G}^{(p)}$  the submatrix of $\mat{G}$ obtained by removing the p-th row and the columns $q \in [\sf K]$ for which $\mat{N}[p,q] = 0$. More formally:
\begin{equation}\label{eqn:sub_G}
	\mat{G}^{(p)} \triangleq  \mat{G}[ p' \neq p,  \{q \colon \mat{N}[p, q]=1\}].
\end{equation}
(Notice the dependence of $\mat{G}^{(p)}$ on the connectivity matrix $\mat{N}$, which is not represented in the notation.)
Let  $g^{(p)}$ be equal to the number of different non-zero integers in the matrix $\mat{G}^{(p)}$.
%\begin{definition}
%	For a given matrix $\mat{M}$, we define $\mathcal{L}_{\mat{M}}$ as the collection of all sets of (0, 1)-matrices such that the average of each set is identical to $\mat{M}$, formally
%	\begin{IEEEeqnarray}{rCl}
%		\mathcal{L}_{\mat{M}} \triangleq \left\{ \mathcal{M} \colon \frac{1}{|\mathcal{M}|} \sum_{\mat{M}' \in \mathcal{M}} \mat{M}' = \mat{M} \right\}
%	\end{IEEEeqnarray} 
%\end{definition}

\begin{theorem}\label{thm:dof_bounds}
	For an ($\mat{M}, \mat{N}$)-channel, the $\SDoF$ of the network defined in Section~\ref{sec:channel_model} is lower bounded by the maximum score $\mathsf{S}$ of our puzzle: 
	\begin{IEEEeqnarray}{rCl} \label{eqn:dof_lower_bound}
		\lefteqn{\SDoF } \nonumber \\
		&\geq &\SDoFlb \triangleq \max_{\mat{G}\in \mathcal{G}_{\mat{M}} } \frac{  \|\mat{G}\|_0}{ \max_{p \in [\sf K]}\left\{ \|\mat{G}[p,:]\|_0 + g^{(p)} \right\}}, \IEEEeqnarraynumspace
	\end{IEEEeqnarray} 
%	where 
%	\begin{IEEEeqnarray}{rCl}
%		T_{\mat{M}'} = \min_{\mat{G} \in \mathcal{G}_{\mat{M}'}}\left\{\max_{p \in [\sf K]} \left\{\|\mat{M'}[p, :]\|_{0} + g^{(p)} \right\}\right\},
%	\end{IEEEeqnarray} 
	where $\mathcal{G}_{\mat{M}}$ is the set of all valid precoding index matrices for the given $\mat{M}$. 
	
\end{theorem}

\begin{IEEEproof}
	See Section \ref{sec:ach}.
\end{IEEEproof}

\subsection{Solution to the puzzle for a class of networks} 
\begin{figure}[h]
 	\centering
 	\includegraphics[width=0.5\textwidth]{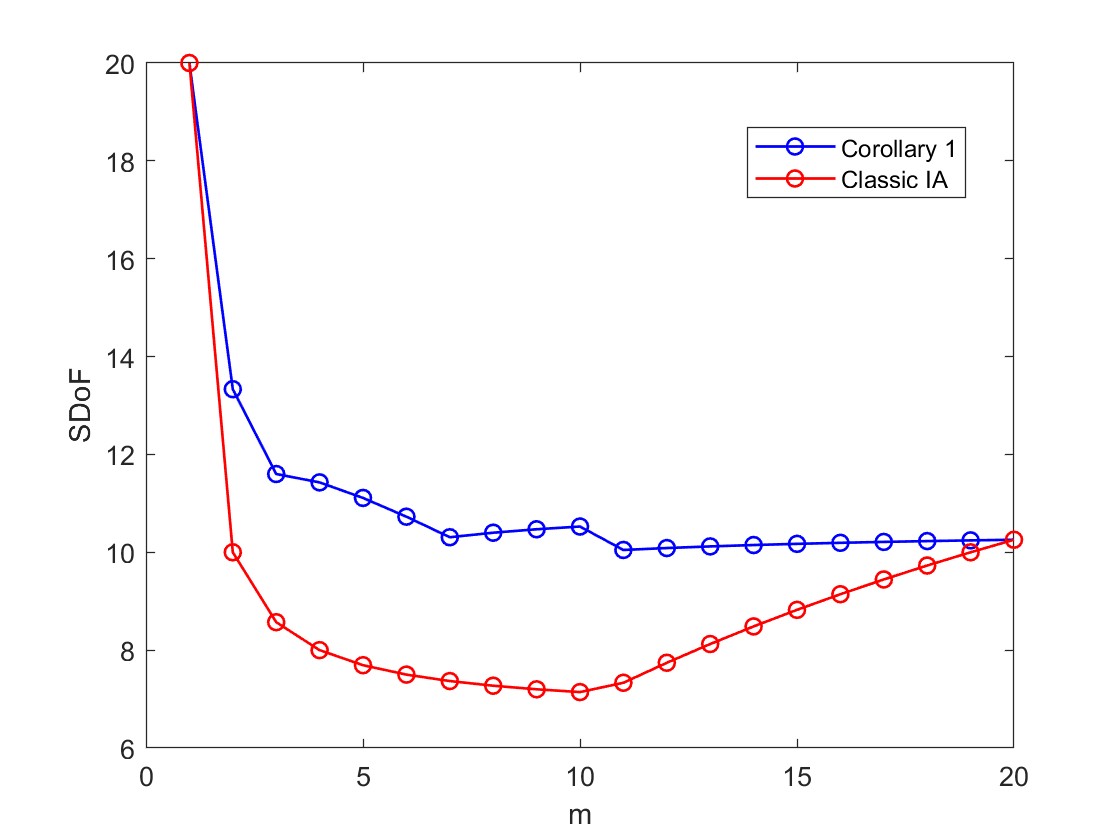}
 	\caption{Comparison the SDoF bounds between Corollary 1 and basic IA scheme for $\sf K=20$}
 	\label{fig:compare}
 	\vspace{-3mm}
 \end{figure}

Consider the special case where Rx $p$ only receives signal  from a Tx $q$ if $(q-p) \mod \sf K < m$, where $m$ is a given positive integer, and each Tx sends an independent message to every connected Rx. In this case, 
\begin{equation}\label{eqn:special_case}
	\mat{M}[p, q] =  \mat{N}[p, q] = \begin{cases}
		1 &\text{ if } (q-p) \mod \sf K < m \\
		0 &\text{otherwise}
	\end{cases}.
\end{equation}
%Notice that each Rx receives $m$ messages, i.e. $\|\mat{M}[p, :]\|_0 = m$, $\forall p \in [\sf K]$. 
\begin{corollary}\label{cor:dof_lower_bounds}
	For the system with matrices $\mat{M}$, $\mat{N}$  in \eqref{eqn:special_case}, we have the following lower bound:
	\begin{equation}
		\SDoFlb = \frac{\sf K \cdot m - (\sf K \mod m)}{2m - 1}.
	\end{equation}
\end{corollary}
\begin{IEEEproof}
	See Section \ref{sec:proof_cor1} for the proof and an example.
\end{IEEEproof}
 Corollary \ref{cor:dof_lower_bounds}  improves the  $\SDoF$ compared to the straightforward implementation of the basic IA scheme, which only achieves a $\SDoF$ of $ \sf K \cdot m / (m + \min\{\sf K-1, 2m-2\})$ (See the appendix for the detailed calculation). We illustrate the two bounds for $\sf K = 20$ in Fig. \ref{fig:compare}.

%\texttt{I think the X-channel is a special case (so an example) of Corollary 1, right? I would simply make it an example. I think the following sentence is sufficient.} 

In the special case $\sf K=m$, our setup coincides with the $X$-channel and our result in Corollary~\ref{cor:dof_lower_bounds} recovers the classic lower bound $\SDoF= \frac{\sf K^2}{2 \sf K -1}$. 

Meanwhile, for the case $\sf m=\sf K-1$, the setup is equivalent to the partially connected X-channel in \cite{isit2022_bi} without cooperation. We recovers the lower bound $\SDoF= \frac{\sf K\cdot(\sf K-1) - 1}{2 \sf K -3}$.

\section{Example of Corollary 1 with $\sf K = 5$ and $m = 2$}
Before discuss the design of matrix $\mat{G}$ for general case, we look at an example with $\sf K = 5$ and $m = 2$. The matrix $\mat{M}$ and $\mat{N}$ are
\begin{equation}
	\mat{M} = \mat{N} =
	\begin{bmatrix}
			1 & 0 & 0 & 0 & 1 \\
			1 & 1 & 0 & 0 & 0 \\
			0 & 1 & 1 & 0 & 0 \\
			0 & 0 & 1 & 1 & 0 \\
			0 & 0 & 0 & 1 & 1
		\end{bmatrix}.
\end{equation}

To obtain $\mat{G}$, we simply replace all $1$ with $2$ for the second and fourth rows of matrix $\mat{M}$. Meanwhile, we fill a $0$ for $\mat{G}[5,5]$. We have
\begin{equation}
	\mat{G} = 
	\begin{bmatrix}
			1 & 0 & 0 & 0 & 1 \\
			2 & 2 & 0 & 0 & 0 \\
			0 & 1 & 1 & 0 & 0 \\
			0 & 0 & 2 & 2 & 0 \\
			0 & 0 & 0 & 1 & \red{0}
		\end{bmatrix}.
\end{equation}
The submatrix $\mat{G}^{(1)}$ is obtained by removing the second, third and fourth columns and the first row of $\mat{G}$ since the second, third and fourth elements in the first row of $\mat{N}$ are zero. 
\begin{equation}
	\mat{G}^{(1)} = 
	\begin{bmatrix}
			2 & 0 \\
			0 & 0 \\
			0 & 0 \\
			0 & \red{0}
		\end{bmatrix},
\end{equation}
and $g^{(1)} = 1$ as there is only one non-zero entry in $\mat{G}^{(1)}$. Similarly, we can deduce that $g^{(2)}$, $g^{(3)}$ and $g^{(4)}$ are also equal to 1. Also $g^{(5)} = 2$ as
\begin{equation}
	\mat{G}^{(5)} = 
	\begin{bmatrix}
			0 & 1 \\
			0 & 0 \\
			0 & 0 \\
			2 & \red{0}
		\end{bmatrix},
\end{equation}
Notice that $\|\mat{G}[p, :]\|_0 = 2$ for $p \in \{1, 2, 3, 4\}$, and $\|\mat{G}[5, :]\|_0 = 1$. Therefore, $\SDoFlb = (2\times5 - 1)/(2+1) = 3$

\lo
{\color{red} I would take this out.
The X channel defined in \cite{jafar_degrees_2008} is a special case of $(\mat{M}, \mat{N})$-channel with $\mat{N}$ and $\mat{M}$ are equal to an all-one matrix, as all receivers can receive signals from all transmitters, and each transmitter sends a message to each receiver. For illustration, we first consider the number of users $\sf K = 5$. We apply the following precoding index matrix
\begin{equation}
	\mat{G} = 
	\begin{bmatrix}
		1 & 1 & 1 & 1 & 1\\
		2 & 2 & 2 & 2 & 2\\
		3 & 3 & 3 & 3 & 3\\
		4 & 4 & 4 & 4 & 4\\
		5 & 5 & 5 & 5 & 5\\
	\end{bmatrix}.
\end{equation}
It is straightforward to observe that the matrix mentioned above is a valid precoding index matrix, as the elements within each column are distinct. Then, the $\mat{G}^{(p)}$ defined in \eqref{eqn:sub_G} is the submatrix obtained by removing the $p$-th row of $\mat{G}$ as $\mat{N}$ is an all-one matrix. As an illustration, when $p=1$, we have
\begin{equation}
	\mat{G}^{(1)} = 
	\begin{bmatrix}
		2 & 2 & 2 & 2 & 2\\
		3 & 3 & 3 & 3 & 3\\
		4 & 4 & 4 & 4 & 4\\
		5 & 5 & 5 & 5 & 5\\
	\end{bmatrix}.
\end{equation}
As there are always four different integers in each submatrix $\mat{G}^{(p)}$, $g^{(p)} = 4$ for all $p \in [5]$. We obtain $\SDoFlb = 25/9$. 

To extend this to any value of $\sf K$, we simply need to choose the precoding index matrix as $\mat{G}[p, q] = p$, $\forall p, q \in [\sf K]$, and this yields the classic result in \cite{jafar_degrees_2008} as the follows:
\begin{equation}
	\SDoFlb = \frac{\sf K^2}{2\sf K - 1}
\end{equation}}

\section{Proof of Corollary \ref{cor:dof_lower_bounds}} \label{sec:proof_cor1}
%Given  given positive integer $\sf K$, $m \in [\sf K]$, and the matrices $\mat{M}$ and $\mat{N}$ defined in \eqref{eqn:special_case}. 
%We focus on the matrix $\mat{M}' \subseteq \mat{M}$ in which the diagonal entries of the last $\sf K \mod m$ rows are changed to zero, i.e. $\mat{M}'[p, p] = 0$ for $p \in \{\lfloor \sf K / m \rfloor m+1, \cdots, \sf K\}$. An example with $\sf K =6$ and $m = 4$ is given below:

The precoding index matrix $\mat{G}$ is constructed by the following steps, where notice that all non-zero elements in $\mat{G}$ are chosen from the set $[m]$.
\begin{enumerate}
\item Initialize all entries of  $\mat{G}$  to zero. 
%\item  Diagonal entries of the last $\sf K \mod m$ rows are changed to zero, i.e. $\mat{M}'[p, p] = 0$ for $p \in \{\lfloor \sf K / m \rfloor+1, \cdots, \sf K\}$.  
	\item \label{enu:step2} For the first $\lfloor \sf K/m \rfloor \cdot m$ rows, fill $p \modd m$ for entries in the $p$-th row if the corresponding entry of $\mat{M}$ is 1. Thus for $p \in [\lfloor \sf K/m \rfloor \cdot m]$ and $q \in [\sf K]$,
	\begin{equation}\label{eqn:G_step2}
		\mat{G}[p, q] = \begin{cases}
			p \modd m &\text{if } \mat{M}[p, q] = 1\\
			0 &\text{if } \mat{M}[p, q] = 0
		\end{cases}
	\end{equation}
	\item \label{enu:step3} For the last  $\sf K \mod m$ rows,  fill all positions where $\mat{M}$ is non-zero with the increasing sequence of numbers  $1, 2, \cdots, m-1$, and set the $p$-th element of the $p$-th row to 0. Thus for $p\in \{\lfloor \sf K / m \rfloor m+1, \cdots, \sf K\}$ and $q \in [\sf K]$:
	\begin{equation}\label{eqn:G_step3}
		\mat{G}[p, q] = \begin{cases}
			i &\text{if }  q = p -m   +i, \ p\neq q\\
			0 &\text{if } \textnormal{otherwise}.
		\end{cases}
	\end{equation}
\end{enumerate}
Notice that the diagonal entries of the last $\sf K \mod m$ rows in $\mat{G}$ are filled with $0$ although those entries in $\mat{M}$ are equal to $1$. We also obtain that 
\begin{equation}\label{eqn:G0}
	\|\mat{G}\|_0 = \sf K \cdot m - (\sf K \mod m)
\end{equation}
%The valid precoding index matrix corresponding to $\mat{M}'$ in \eqref{eqn:example_M64} is
For example, for $\sf K=6$ and $m=4$, we have:
\begin{equation}\label{eqn:example_M64}
	\mat{N}=\mat{M}
	\begin{bmatrix}
		1 & 0 & 0 & 1 & 1 & 1\\
		1 & 1 & 0 & 0 & 1 & 1\\
		1 & 1 & 1 & 0 & 0 & 1\\
		1 & 1 & 1 & 1 & 0 & 0\\
		0 & 1 & 1 & 1 & 1 & 0\\
		0 & 0 & 1 & 1 & 1 & 1\\
	\end{bmatrix},	
\end{equation}
and
\begin{equation}\label{eqn:example_M64_G}
	\mat{G} = 
	\begin{bmatrix}
		1 & 0 & 0 & 1 & 1 & 1\\
		2 & 2 & 0 & 0 & 2 & 2\\
		3 & 3 & 3 & 0 & 0 & 3\\
		4 & 4 & 4 & 4 & 0 & 0\\
		0 & \bl {1} & \bl{2} & \bl{3} & \red{0} & 0\\
		0 & 0 & \bl{1} & \bl{2} & \bl{3} & \red{0}\\
	\end{bmatrix}
\end{equation}
where the first four rows are filled by following step \ref{enu:step2}, the last two rows (marked in blue) are filled by following step \ref{enu:step3}, and notice that the diagonal entries of the last two rows remains $0$ (marked in red).

The proposed  matrix $\mat{G}$ is a valid precoding index matrix because:
\begin{itemize} 
\item $\mat{G}[p, q]>0$ only when $\mat{M}[p,q]>0$. 
\item Each positive number occurs only once in a row. To see this, focus on the integer $i$. It occurs  $m$ times  in row $i$, at columns $i, i-1, i-2, \ldots, i-m+1$, where differences have to be understood $\modd \sf K$, and if $i \neq m$ it also occurs once in each of the  rows $p \in   \{ \lfloor \sf K / m \rfloor m+1, \cdots, \sf K\}$ at  column $p -m  +i$. %, where this difference  again has to be understood modulo $\sf K$.
Symbol $i<m$ thus occurs in columns (up to  $\modd \sf K$) $i, i-1, \ldots, i- m +1,  p - m +1, \ldots,  p -1 $, which are all distinct numbers. Symbol $m$ only occurs in row $m$ and thus at most once per column. 
\end{itemize}  %positive integers for each entry where the corresponding entry in $\mat{M}'$ is one. For the second requirement, as the whole system is symmetric with respect to all non-zeros integers, we focus on $1$ in $\mat{G}$. Notice that, in the $p$-th row with $p \in [\lfloor \sf K/m \rfloor \cdot m]$, the non-zeros elements in $\mat{M}'$ are $\{\mat{M}'[p, p-m+1], \mat{M}'[p, p-m+2], \cdots, \mat{M}'[p, p]\}$. Combining with \eqref{eqn:G_step2}, we find that $1$ presents once in each of the columns 
%$ 1, 2, \cdots, (\lfloor \sf K/m\rfloor-1)m \text{ and } \sf K.  $ Consider \eqref{eqn:G_step3}, we find $1$ presents once in each of the columns 
%$ (\lfloor \sf K/m\rfloor-1)m+1, \cdots, \sf K -1.  $
%Therefore, 1 presents in each column in $\mat{G}$ without repetition, which implies $\mat{G}$ is a valid precoding index matrix. 

We now compute the value of $g^{(p)}$ for all indices $p\in[\sf K]$, for which we need submatrices $\mat{G}^{(p)}$.  As an example, consider  the  two submatrices   $	\mat{G}^{(1)}$ and $	\mat{G}^{(5)}$ extracted from \eqref{eqn:example_M64_G}:
\begin{equation}
	\mat{G}^{(1)}=
	\begin{bmatrix}
	2 & 0 & 2 & 2\\
	3 & 0 & 0 & 3\\
	4 & 4 & 0 & 0\\
	0 & \bl{3} & \red{0} & 0\\
	0 & \bl{2} & \bl{3} & \red{0}\\
	\end{bmatrix}, 
	\mat{G}^{(5)}=
	\begin{bmatrix}
	0 & 0 & 1 & 1 \\
	2 & 0 & 0 & 2 \\
	3 & 3 & 0 & 0 \\
	4 & 4 & 4 & 0 \\
	0 & \bl{1} & \bl{2} & \bl{3} \\
	\end{bmatrix}
\end{equation}

For   $p \in \{\lfloor \sf K/m\rfloor\cdot m, \cdots, \sf K\}$ we have the trivial value $g^{(p)} = m$.  because  all elements are present in the submatrix. However, for 
 $p \in [\lfloor \sf K/m\rfloor\cdot m]$ the element  $p \modd m$ is not contained in the submatrix $\mat{G}^{(p)}$ and thus  $g^{(p)} = m-1$.    Notice next that with above algorithm  $\|\mat{G}[p, :]\|_0 = m$ for $p \in [\lfloor \sf K/m\rfloor\cdot m]$, and $\|\mat{G}[p, :]\|_0 = m-1$ for $p \in \{\lfloor \sf K/m\rfloor\cdot m, \cdots, \sf K\}$. Therefore,  we have the constant value
 \begin{equation}\label{eqn:T_GN}
 \|\mat{G}[p, :]\|_0 + g^{(p)} = 2m-1, \quad \forall p \in [\sf K],
 \end{equation} establishing the desired lower bound for $\SDoF$ by calculate the ratio between \eqref{eqn:G0} and \eqref{eqn:T_GN}.

\section{Proof of the \SDoF  Lower Bound in Theorem~\ref{thm:dof_bounds}}\label{sec:ach}

\subsubsection{Coding Scheme}
We fix a parameter $\eta \in \mathbb{Z}^+$
%\yue{\begin{IEEEeqnarray}{rCl}
		%\Gamma  \triangleq  \sf K \cdot (\sf K-2) \cdot \sf r^2, 
		%\end{IEEEeqnarray}}
		and a valid precoding index matrix $\mat{G}$. Define
		\begin{IEEEeqnarray}{rCl}
			p_{max} =  \argmax_{p \in [\sf K]} \left\{ \|\mat{G}[p, :]\|_0 + g^{(p)} \right\}.
		\end{IEEEeqnarray}
		choose 
		\begin{IEEEeqnarray}{rCl}
			\sf T =   \eta^\Gamma \|\mat{G}[p_{max}, :]\|_0 + (\eta+1)^\Gamma g^{(p_{max})} .
		\end{IEEEeqnarray}
		where we specify the value of $\Gamma$ later in Eq. \eqref{eqn:Gamma}.
		%Split the block length $\sf T$ into $\sf T''\triangleq \sf T/\mu$ blocks of length $\mu$.
		We only send messages $\{a_{p,q}\}$ with $\mat{G}[p,q]>0$. Each message is encoded using a circularly symmetric Gaussian codebook of average power $\sf P / \|\mat{G}[:, q]\|_0$ and  codeword  length $\eta^\Gamma$. Each codeword is sent over a block of $\sf T$ consecutive channel uses.
		More precisely, let $\{\vect{b}_{p,q}\}$ denote the $\eta^\Gamma$-length codeword symbol for message $a_{p,q}$. Each Tx $q$ form their inputs as:
		\begin{IEEEeqnarray}{rCl}
			\vect{X}_q &=& \sum_{\{p \colon \mat{G}[p,q]>0\}} \mat{U}_{\mat{G}[p,q]} \vect{b}_{p,q}, \label{eq:X}
		\end{IEEEeqnarray}
		where matrices $\{\mat{U}_g\}_{g \in [g_{max}]}$ are described shortly, and $g_{max}$ is the maximal element in $\mat{G}$. Notice that messages $a_{p, q}$ and $a_{p', q'}$ are actually multiplied by the same precoding matrix if $\mat{G}[p, q] =\mat{G}[p', q'] > 0$.
		
		We recall that Rx $p$ only receives signals from connected Txs, i.e. from Tx $q$ with $\mat{N}[p, q]=1$, which allows to write the observed signal at each Rx $p \in [\sf K]$ as:
		\begin{IEEEeqnarray}{rCl}\label{eq:Yp}
			\vect{Y}_p
			&=&\underbrace{ \sum_{\{q \colon \mat{G}[p,q]>0\}} \mat{H}_{p,q} \mat{U}_{\mat{G}[p,q]}\vect{b}_{p,q} }_{\text{desired signal}} \nonumber\\
			& + &\underbrace{ \sum_{\{q \colon \mat{N}[p,q]=1\}} \sum_{ \substack { \{p' \colon  p' \neq p\\ \mat{G}[p', q] > 0 \}}} \mat{H}_{p,q} \mat{U}_{\mat{G}[p',q]}\vect{b}_{p',q} }_{\text{Interference}} + \vect{Z}_p,\nonumber\\[-2ex]
		\end{IEEEeqnarray}
		where
			$\mat{H}_{p,q} \triangleq \text{diag}(\left[H_{p,q}(1), H_{p,q}(2) \cdots H_{p,q}(\sf T)\right]),$ 
		$\vect{Y}_p \triangleq (Y_{p}(1),\ldots,Y_{p}(\sf T) )^\intercal,$ and $\vect{Z}_p$ are the corresponding Gaussian noise vectors observed at Rx $p$. 
		
		\subsubsection{IA Matrices $\{\mat{U}_{g}\}$} 
		Inspired by the IA scheme in \cite{jafar_degrees_2008}, we choose each $\sf T \times \eta^\Gamma$ precoding matrix $\mat{U}_g$ so that its column-span includes all power products (with powers from 1 to $\eta$) of the channel matrices $\mat{H}_{p, q}$ that premultiply $\mat{U}_g$ in \eqref{eq:Yp} when the coded symbol is treated as interference for the receiver. That implies, for $g \in [g_{max}]$:
		\begin{IEEEeqnarray}{rCl}\label{eq:U_i}
			\mat{U}_{g}= \left[ \prod_{\mat{H} \in {\mathcal{H}}_{g} }
			\mat{H}^{{\alpha_{g,\mat{H}}}}\cdot  \boldsymbol{\Xi}_{g}\colon  \, \; 
			\forall \boldsymbol{\alpha}_g \in  [\eta]^{\Gamma} \right], \quad 
		\end{IEEEeqnarray}
		where 
		$\{\boldsymbol{\Xi}_g\}_{g \in \mat{G}}$ are i.i.d. random vectors independent of all channel matrices, noises, and messages,  
		\begin{IEEEeqnarray}{rCl}
			\label{eq:set_H}
			\mathcal{H}_g &\triangleq& \left\{  \mat{H}_{p, q} \colon \mat{N}[p, q] = 1 \text{ and } \exists p' \in [\sf K]\backslash \{p\}\right. \nonumber \\ && \left.\text{ such that } \mat{G}[p', q]=g \right\} , 
		\end{IEEEeqnarray}
		$\boldsymbol{\alpha}_g \triangleq (\alpha_{g, \mat{H}}\colon \quad \mat{H} \in {\mathcal{H}}_{g})$, and we assume the sizes of $\mathcal{H}_g$ for $g \in [g_{max}]$ are equal \footnote{If $|\mathcal{H}_g|$ varies for different $g$, additional i.i.d. diagonal random matrices can be added to $\mathcal{H}_g$ with fewer element, and the rest of proof remains unchanged.} and \begin{equation}\label{eqn:Gamma}
				\Gamma = |\mathcal{H}_g|.
			\end{equation}

		\subsubsection{Analysis of Signal-and-Interference Subspaces}
		Since the column-span of $\mat{U}_g$ contains all power products  of powers 1 to $\eta$ of the channel matrices $\mat{H}_{p,q}$ that premultiply $\mat{U}_g$ in \eqref{eq:Yp}, the product of any of these matrices with $\mat{U}_g$ is included in the column-space of the $\sf T\times  (\eta+1)^\Gamma$-matrix
		\begin{IEEEeqnarray}{rCl}\label{eq:W_i}
			\mat{W}_{g} =  \left[ \prod_{\mat{H} \in {\mathcal{H}}_{g} }
			\mat{H}^{{\alpha_{g,\mat{H}}}} \cdot \boldsymbol{\Xi}_{g}\colon  \; \, 
			\forall \boldsymbol{\alpha}_g \in  [\eta+1]^{\Gamma} \right]&, \nonumber\\
			g \in [g_{max}].
			%\textnormal{ for } i\in[ \tilde{\sf K}]\backslash\{1\}&,
		\end{IEEEeqnarray}
		Formally, for each $g \in [g_{max}]$ and $\mat{H} \in \mathcal{H}_g$, we have $\text{span}(\mat{H}\cdot \mat{U}_g) \subseteq \text{span}(\mat{W}_g)$. 
		%Noting also that by the  block-diagonal structures of $\{\mat{S}^{(j,k)}\}$ and $\{\tilde{\mat{U}}_i\}$, the signal space at any single Receiver $j\in[K]$ is given by $[\mat{S}^{(1,2)} \mat{U}_2 ]$ for Receivers $j=1,...,$ and by $...$ for Receivers $j=1,...$. 
		As a consequence, the signal and interference space at a Rx~$p\in [\sf K]$ is represented by the % $\mu \times ((\tilde{\sf K}-1)\eta^{\Gamma} + (\tilde{\sf K}-2)\eta^{(\Gamma+1)})$
		matrix: 
		\begin{IEEEeqnarray}{rCl}\label{eq:signal_interference2}
			\boldsymbol{\Lambda}_p \triangleq  \left[ \mat{D}_p, \mat{I}_p \right] .  \IEEEeqnarraynumspace
		\end{IEEEeqnarray}
		with the signal subspaces given by the $\sf T \times (\tilde{\sf K}-1)\eta^{\Gamma}$-matrices
		\begin{IEEEeqnarray}{rCl}
			\mat{D}_p  \triangleq \left[ \mat{H}_{p,q}\mat{U}_{\mat{G}[p, q]} \colon \mat{G}[p,q]>0 \right].
		\end{IEEEeqnarray}
		and the interference subspaces given by 
		\begin{IEEEeqnarray}{rCl}
			\mat{I}_p \triangleq \left[ \mat{W}_{\mat{G}[p',q]}\colon \mat{N}[p,q]=1, p' \neq p, \mat{G}[p', q]>0 \right]
		\end{IEEEeqnarray}
		
		%For a Rx $p$ in the first group $\TS{1}$, the signal and interference spaces are represented by the $\sf T \times \sf T$-matrix:
		%\begin{IEEEeqnarray}{l}\label{eq:signal_interference1}
		%\boldsymbol{\Lambda}_p=
		%\big[ \underbrace{\mat{D}_{p,2}, \; \cdots,  \; \mat{D}_{p,\tilde{\sf K}-1},}_{\textnormal{signal space}}  \;  \underbrace{{\mat{W}}_{2},\; {\mat{W}}_{3} , \;\cdots, \;{\mat{W}}_{\tilde{\sf K}}}_{\textnormal{interference space}} \big], \IEEEeqnarraynumspace
		%\end{IEEEeqnarray}
		%where the signal subspace is given by the $\sf T \times \eta^{\Gamma}$-matrices
		%\begin{IEEEeqnarray*}{rCl}
		%    \mat{D}_{p,k} \triangleq \mat{S}^{(1,k)}_p \cdot \mat{U}_k, \quad  k \in \{2,...,\tilde{\sf K}-1\} ,\quad  p \in \TS{1}.
		%\end{IEEEeqnarray*}
		By observing the signal subspace, we obtain the following property
		\begin{property}\label{pro:signal_subspace_index}
			For matrix $\mat{H}_{p,q} \mat{U}_{\mat{G}[p,q]}$ in the signal subspace $\mat{D}_p$, we have $\mat{H}_{p,q} \notin \mathcal{H}_{G[p,q]}$ according to the second condition on the valid precoding index matrix.
		\end{property}
		\begin{IEEEproof}
			The property can be proven by contradiction. We assume $\mat{H}_{p,q} \in \mathcal{H}_{G[p,q]}$. By the definition of $\mathcal{H}$ in \eqref{eq:set_H}, we deduce that $\exists p' \in [\sf K]\backslash \{p\}$ such that $\mat{G}[p',q] = \mat{G}[p, q]$. There is thus at least one non-zero integer repeats within a column of $\mat{G}$, which violates the second requirement of a valid precoding index matrix in Definition \ref{def:valide_precoding_index_matrix}. This concludes the proof.
		\end{IEEEproof}
		
		We shall prove that  all matrices $\{\mat{\Lambda}_p\}$ are of full column rank. This proves  that the desired signals  intended for Rx $p$ can be separated from each other and from the 
		interference space at this Rx. In the limits $\eta \to\infty$ (and thus $\sf T \to \infty$) and $\sf P\to\infty$, this establishes an DoF of  
		$$\lim_{\eta \to \infty} \frac{\|\mat{G}[p, :]\|_0 \eta^{\Gamma}}{\sf T} = \frac{ \|\mat{G}[p, :]\|_0 }{ \|\mat{G}[p_{max}, :]\|_0 + g^{(p_{max})}}$$
		at Rxs $p$. 
		\lo{The \yue{\SDoF} is therefore given by
			\begin{IEEEeqnarray}{rCl}
				\yue{\SDoF} &=& \sf r \cdot \left(\frac{ \tilde{\sf K}-1 }{ 2 \tilde{\sf K}-3} \cdot (\tilde{\sf K}-1)+ \frac{ \tilde{\sf K}-2 }{ 2 \tilde{\sf K}-3}\right) \\
				&=& \frac{\sf K\cdot (\sf K - \sf r) -\sf r^2}{2\sf K - 3\sf r} = \yue{\SDoFlb},
			\end{IEEEeqnarray}
		}
		\sh{The $\SDoF$ of the entire system is thus given by $\SDoFlb$,} which  establishes the desired result.
		In the following, we will explain in detail that the matrix $\mat{\Lambda}_p$ has the same form as the matrix $\mat{A}$ in Lemma \ref{lma:lemma1} and satisfy the two conditions mentioned in Lemma \ref{lma:lemma1}. Then any square submatrix of $\mat{\Lambda}_p$ has the same form as $\mat{A}$, which by Lemma \ref{lma:lemma1} proves that the matrix $\mat{\Lambda}_p$ is full column rank with probability 1.
		
		To see that $\mat{\Lambda}_p$ is of the form in \eqref{eq:lemma1_A}, notice that all matrices involved in \eqref{eq:signal_interference2}, i.e. $\{\mat{H}_{p,q}\}$ ,  are diagonal, and their multiplications with an vector, i.e. $\{\boldsymbol{\Xi}_{g}\}$, from the right leads to a column-vector consisting of the non-zero entries of these diagonal matrices. More precisely, the random variables in row $t$ are given by the slot-$t$ channel coefficients $\{H_{p,q}(t)\}$ and the $t$-th elements of vector $\{\boldsymbol{\Xi}_{g}\}$, which by definition are independent of each other and of all random variables in the other rows. Therefore, the matrix $\mat{\Lambda}_p$ satisfies Condition~i) in Lemma~\ref{lma:lemma1}. To see that it also satisfies Condition~ii), notice that for any two distinct columns $\vect{v}^{(1)}$ and $\vect{v}^{(2)}$ selected from $\mat{\Lambda}_p$, the exponents in the corresponding columns differ because:
		\begin{enumerate}
			\item If $\vect{v}^{(1)}$ and $\vect{v}^{(2)}$ are selected from the same signal subspace $\mat{H}_{p,q}\mat{U}_{g}$ or the same interference subspace $\mat{W}_{g}$, the two vectors have different exponents due to the construction method of $\mat{U}$ and $\mat{W}$.
			\item If $\vect{v}^{(1)}$ is selected from the signal subspace $\mat{H}_{p,q}\mat{U}_{g}$, and $\vect{v}^{(2)}$ is selected from the signal subspace $\mat{H}_{p,q'}\mat{U}_{g'}$ or interference subspace $\mat{W}_{g'}$ with $g \neq g'$, the two vectors have different exponents as they have distinct factors $\boldsymbol{\Xi}_{g}$ and $\boldsymbol{\Xi}_{g'}$.
			\item If $\vect{v}^{(1)}$ is selected from the signal subsapce $\mat{H}_{p,q}\mat{U}_{g}$, and $\vect{v}^{(2)}$ is selected from the signal subspace $\mat{H}_{p,q'}\mat{U}_{g}$ or interference subsapce $\mat{W}_{g}$, we can deduce that $\mat{H}_{p,q} \notin \mathcal{H}_g$ by Property \ref{pro:signal_subspace_index}. $\vect{v}^{(1)}$ has the factor $\mat{H}_{p,q}$ while $\vect{v}^{(2)}$ does not. The two vectors thus have different exponents.
		\end{enumerate}
		This concludes the proof. 
		\begin{lemma}[Lemma~1 in \cite{cadambe_interference_2009}] \label{lma:lemma1}
			Consider an $\sf M$-by-$\sf M$ square matrix $\textbf{A}$ with $i$-th row and $j$-th column entry \begin{equation} \label{eq:lemma1_A}
				a_{ij} = \prod_{\ell=1}^{\sf L} \left( X_i^{[\ell]}\right)^{\alpha_{ij}^{[\ell]}}, \qquad  i,j \in \sf M,
			\end{equation}
			for  random variables  $\{X_i^{[\ell]}\}_{\ell \in [\sf L]}$ and exponents 
			\begin{equation}
				{\boldsymbol\alpha}_{ij} \triangleq \left(\alpha_{ij}^{[1]}, \alpha_{ij}^{[2]}, \ldots, \alpha_{ij}^{[\sf L ]}\right) \in \mathbb{Z}^{+\sf L}.
			\end{equation} 
			If 
			\begin{enumerate}
				\item for any two pairs $(i,\ell) \neq (i', \ell')$ the conditional cumulative probability distribution $P_{ X_i^{[\ell]}|X_{i'}^{[\ell']}}$ is continuous; and
				\item any pair of vectors ${\boldsymbol\alpha}_{i,j} \neq {\boldsymbol\alpha}_{i,j'}$ for $i, j,j' \in [\sf M]$ with $j \neq j'$;
			\end{enumerate}
			then the matrix $\textbf{A}$ is full rank with probability 1.
		\end{lemma}

\section{Conclusion}\label{sec:ccl}
In conclusion, our study revolves around a $\sf K$-user channel characterized by a connectivity matrix $\mat{N}$ and a message  matrix $\mat{M}$. To overcome this limitation and facilitate the design of new IA schemes, we introduce a formalized number-filling puzzle. Our analysis establishes a compelling relationship: the score of a puzzle solution is  proportional to the sum degrees of freedom achieved by the corresponding IA scheme, given any $\mat{M}$ and $\mat{N}$. Leveraging this puzzle, we uncover new IA schemes for symmetric channels, demonstrating enhanced $\SDoF$ for sparse connectivity channels compared to the straightforward implementation of the basic IA scheme.

\section*{Acknowledgement}
The authors would like to thank Michèle Wigger and Pilippe Ciblat for helpful discussions and inspirations for this work. This work has been supported by National Key R\&D Program of China under Grant No 2020YFB1807504 and National Science Foundation of China Key Project under Grant No 61831007.

\appendices
\section{Comparison the SDoF bounds between Corollary 1 and basic IA scheme}
When applying the basic IA scheme, we also assign the same precoding matrix $\mat{U}_p$ for messages intended to Rx $p$. Thus, for given $\sf K$, $m$, $\mat{M}$ and $\mat{N}$, the precoding index matrix is 
\begin{equation}
	\mat{G}[p,q] =
	\begin{cases}
		p &\text{if } \mat{M}[p,q]=1\\
		0 &\text{if } \mat{M}[p,q]=0.
	\end{cases}
\end{equation}
We revisit the same example for $\sf K = 5$ and $m = 2$. We have 
\begin{equation}
	\mat{M} = \mat{N} =
	\begin{bmatrix}
		1 & 0 & 0 & 0 & 1 \\
		1 & 1 & 0 & 0 & 0 \\
		0 & 1 & 1 & 0 & 0 \\
		0 & 0 & 1 & 1 & 0 \\
		0 & 0 & 0 & 1 & 1
	\end{bmatrix}.
\end{equation}
and 
\begin{equation}
	\mat{G} = 
	\begin{bmatrix}
		1 & 0 & 0 & 0 & 1 \\
		2 & 2 & 0 & 0 & 0 \\
		0 & 3 & 3 & 0 & 0 \\
		0 & 0 & 4 & 4 & 0 \\
		0 & 0 & 0 & 5 & 5
	\end{bmatrix}.
\end{equation}
Now, we need to determine $g^{(p)}$ for $p \in[\sf K]$. As the matrix $\mat{G}$ is symmetric in terms of elements, we focus on the submatrix $\mat{G}^{(1)}$ and $g^{(1)}$. It is straightforward to find that $g^{(1)} = 3$ as
\begin{equation}
	\mat{G}^{(1)} = 
	\begin{bmatrix}
		2 &  0 \\
		0 &  0 \\
		0 &  0 \\
		0 &  5
	\end{bmatrix}.
\end{equation}
For general cases, in the first row of $\mat{N}$, we have
\begin{equation}\label{eqn:non-zeros_columns}
	\mat{N}[1, 2] = \mat{N}[1, 3] =  \cdots = \mat{N}[1, \sf K-m+1] = 0,
\end{equation}
which are $\sf K - m$ adjacent entries. $\mat{G}^{(1)}$ does not contain the corresponding columns $2, 3, \cdots \sf K-m+1$ in $\mat{G}$. We observe that the $p$-th row in $\mat{M}$ can be obtained by right shifting the $p-1$-th row by one position. Therefore, there are $\max\{0, \sf K - 2m +1\}$ rows in $\mat{G}$ (except the first row) whose all non-zeros entries are in columns mentioned in \eqref{eqn:non-zeros_columns}, i.e. $2, 3, \cdots \sf K-m+1$. Notice that integers are different between rows. We obtain $g^{(1)} = \min\{\sf K - 1, 2m-2\}$. The $\SDoF$ achieved is 
\begin{equation}\label{eqn:naive_SDoF}
	\textnormal{Sum-DoF}_\textnormal{classic} =
	\frac{\sf K \cdot m}{m + \min\{\sf K - 1, 2m-2\}}
\end{equation}

To compare the two bounds, we examine the sign of $\SDoFlb - \textnormal{Sum-DoF}_\textnormal{classic}$. For $2m \leq \sf K$, we have $\sf K \mod m > m-1$, and we only need to examine the sign of
\begin{equation}
	(\sf K m - m + 1)(3m-2) - \sf K m(2m-1) = (\sf Km -m+1)(m-1) > 0
\end{equation}
For $2m > \sf K$, we have $\sf K \mod m = K - m$, and we only need to examine the sign of
\begin{equation}
	(\sf K -\sf K +m)(m+\sf K -1) - \sf Km(2m - 1) = (\sf K-m)(\sf K-1)(m-1) \geq 0.
\end{equation}
Therefore, we always have
\begin{equation}
	\SDoFlb \geq \textnormal{Sum-DoF}_\textnormal{classic}
\end{equation}

\bibliographystyle{IEEEtran}
\bibliography{IEEEabrv,references}

\end{document}